\documentclass[runningheads]{llncs}
\usepackage[T1]{fontenc}
\usepackage{graphicx}
\usepackage{nicefrac}
\usepackage{subfig}
\usepackage{multirow}
\usepackage{bbold}
\usepackage{balance}
\usepackage{enumitem}

\usepackage[hidelinks]{hyperref}
\PassOptionsToPackage{hyphens}{url}
\usepackage{url}
\usepackage{hyperref}
\hypersetup{
  colorlinks   = true, %Colours links instead of ugly boxes
  urlcolor     = blue, %Colour for external hyperlinks
  linkcolor    = black, %Colour of internal links
  citecolor   = blue %Colour of citations
}

\usepackage{footnote}
\usepackage{epsfig}
\usepackage{amsmath}
\usepackage{amsfonts}
\usepackage{relsize}

\usepackage{stackengine}
\usepackage{algorithmic,comment}
\usepackage[ruled,vlined,commentsnumbered,titlenotnumbered,linesnumbered]{algorithm2e}
\usepackage{float}

\usepackage[font=small,labelfont=bf]{caption}

\usepackage{colortbl}
\usepackage{url}
\usepackage{color}

\newcommand{\numconsumer}{n_t}

\newcommand{\totalsupply}{S_t}
\newcommand{\tokenreserve}{R^{\text{Tok}}_t}
\newcommand{\dollarreserve}{R^{\text{USD}}_t}
\newcommand{\dollarreservemin}{R^{\text{USD,min}}_t}
\newcommand{\buybackdollars}{u^{\text{B}}_t}
\newcommand{\paytokens}{u^{\text{P}}_t}
\newcommand{\demand}{D_t}
\newcommand{\tokenprice}{p^{\text{Tok}}_t}
\newcommand{\bartokenprice}{\bar{p}^{\text{Tok, ref.}}_t}

\newcommand{\reserveincomes}{\text{Inc}_t}

\newcommand{\incentive}{\Delta p_t}
\newcommand{\buybacktokens}{\frac{u^{\text{B}}_t}{\tokenprice + \incentive}}
\newcommand{\state}[1]{x_{#1}}
\newcommand{\control}[1]{u_{#1}}

\newcommand{\forecasted}[1]{s_{#1}}

\newcommand{\utility}{U_t(\alpha_t,\incentive)}

\newcommand{\reals}{\mathbb{R}}

\begin{document}
\title{A Control Theoretic Approach to Infrastructure-Centric Blockchain Tokenomics}
\titlerunning{A Control Theoretic Approach to Blockchain Tokenomics}
%\title{A Systems Theoretic Approach to Infrastructure-Centric Blockchain Tokenomics}
%\titlerunning{A Systems Theoretic Approach for Blockchain Tokenomics}

%\author{First Author\inst{1} \and
%Second Author\inst{2,3}\orcidID{1111-2222-3333-4444} \and
%Third Author\inst{3}\orcidID{2222--3333-4444-5555}}
%% First names are abbreviated in the running head.
%\institute{Princeton University, Princeton NJ 08544, USA \and
%Springer Heidelberg, Tiergartenstr. 17, 69121 Heidelberg, Germany

%\clearheadinfo
\author{Oguzhan Akcin, Robert P. Streit, Benjamin Oommen, \\ Sriram Vishwanath, and Sandeep Chinchali}

\institute{The University of Texas at Austin \email{\{oguzhanakcin, rpstreit, \\baoommen, sriram, sandeepc\}@utexas.edu}
}

\authorrunning{ O. Akcin et al. }

\maketitle              % typeset the header of the contribution
\begin{abstract}
There are a multitude of Blockchain-based physical infrastructure systems, ranging from decentralized 5G wireless to electric vehicle charging networks. 
These systems operate on a crypto-currency enabled token economy, where node suppliers are rewarded with tokens for enabling, validating, managing and/or securing the system. However, today's token economies are largely designed without infrastructure systems in mind, and often operate with a fixed token supply (e.g., Bitcoin). Such fixed supply systems often encourage early adopters to hoard valuable tokens, thereby resulting in reduced incentives for  new nodes when joining or maintaining the network. 
This paper argues that token economies for infrastructure networks should be structured differently -- they should 
\textit{continually incentivize} new suppliers to join the network to provide services and support to the ecosystem.
%We argue that a token economy for infrastructure networks should \textit{continually incentivize} new suppliers to join the blockchain and provide ubiquitous services, such as 5G connectivity. 
As such, the associated token rewards should gracefully scale with the size of the decentralized system, but should be carefully balanced with consumer demand to manage inflation and be designed to ultimately reach an equilibrium. To achieve such an equilibrium, the decentralized token economy should be \textit{adaptable} and controllable so that it maximizes the total utility of all users, such as achieving stable (overall non-inflationary) token economies.

%\hspace{5pt} 
\-\hspace{11pt}
Our main contribution is to model infrastructure token economies as \textit{dynamical systems} -- 
the circulating token supply, price, and consumer demand change as a function of the payment to nodes and costs to consumers for infrastructure services. Crucially, this dynamical systems view enables us to leverage  tools from mathematical control theory to optimize the overall decentralized network’s performance.
Moreover, our model extends easily to a Stackelberg game between the controller and the nodes, which we use for robust, strategic pricing.
%to bolster our techniques against rational behavior. 
In short, we develop predictive, optimization-based controllers that outperform traditional algorithmic stablecoin heuristics by up to $2.4 \times$ in simulations based on real demand data from existing  decentralized wireless networks.

\end{abstract}

\section{Introduction}

The space of Blockchain-based physical infrastructure networks is rapidly growing, including  decentralized wireless, storage, compute, and electric vehicle charging networks. As an example, Helium \cite{HeliumWhitePaper} and Pollen \cite{PollenWhitePaper} are two prominent decentralized wireless networks (DeWi) that reward the general public to build, maintain, validate, secure and ultimately, send data over 5G hotspots. Similarly, projects such as FileCoin \cite{FilecoinWhitePaper}, Storj
\cite{StorjWhitePaper} and ComputeCoin \cite{ComputeCoinWhitePaper} offer decentralized file storage and computing services. These networks reward suppliers using a corresponding (cryptocurrency) token to build, maintain, secure, and offer  services over this decentralized infrastructure network. Likewise, consumers can often exchange US dollars (USD) for tokens, which enables them to utilize infrastructure services and/or participate in the associated crypto-economy. 

Despite the popularity of decentralized infrastructure networks, we lack systematic tools to design their token economies to incentivize supply growth and consumer demand. Today’s token economies largely target finance, such as Bitcoin, and can operate with a (typically) fixed supply of tokens. However, these fixed supply monetary systems are starkly different from physical infrastructure networks. For example, in a fixed supply system such as Bitcoin, early adopters can hoard tokens since
they are scarce. Moreover, late adopters might not be adequately incentivized to join or maintain the network as token rewards could prove to be smaller than those of early participants. 

\begin{figure}[t]
    \begin{minipage}[c]{0.6\textwidth}
    \includegraphics[width=1.0\columnwidth]{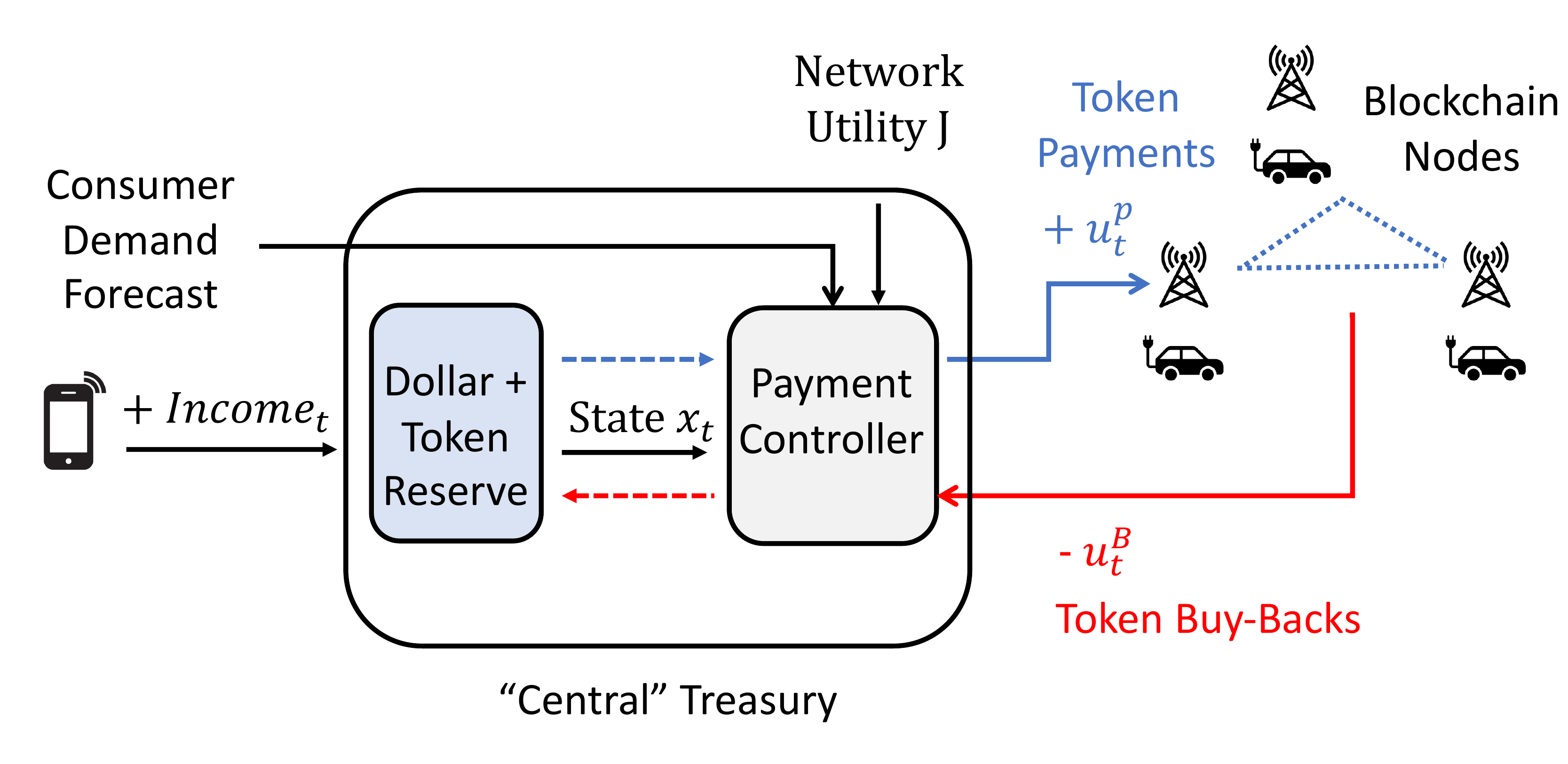}
      \end{minipage}\hfill
      \begin{minipage}[c]{0.38\textwidth}
\caption{\small{\textbf{A Control System for Blockchain Tokenomics:}
    We design a controller (gray) to achieve a burn and mint equilibrium. The controller takes in a forecast of consumer demand and supplier growth $s_t$, as well as the treasury state $x_t$. Then, it adaptively controls token payments $u^P_t$ and buy-backs $u^B_t$ to achieve a stable token price.
        \label{fig:tokenomics}
    } }
    \end{minipage} 
\vspace{-1.5em}
\end{figure}

Our central thesis is that a token economy must be designed to continually incentivize new suppliers to join the ecosystem and provide  services, such as 5G connectivity for Helium or electric vehicle charging stations. As such, the number of tokens should gracefully scale with the size of the infrastructure network, which we do not know a-priori. However, continually rewarding suppliers with newly created tokens can result in inflation if such payments are not carefully balanced
with the consumer demand for infrastructure services. To solve such problems, a number of projects have recently considered adopting/adopted a ``burn-and-mint'' token economics (tokenomics) model, where a central reserve ``mints'' tokens to reward suppliers, while tokens are ``burnt'' (deleted from the circulating supply) when consumers want to use network services. By adaptively burning tokens, we can reduce the token supply to reach an overall supply-demand equilibrium. 

Moreover, such a burn-and-mint equilibrium (BME) \cite{messaricryptonews} must  be ``programmable'' so that  Blockchain-based infrastructure networks can maximize the total utility of all users. For example, this network utility function (performance criterion) can include maintaining a stable, steadily growing token price with low volatility. Likewise, this network cost function can incentivize new suppliers/consumers to expand geographical coverage. Moreover, the BME-based token economy could be designed to satisfy
strict performance guarantees and constraints, such as limiting the number of tokens minted and/or burned per day.
Taking this even further, participants in the economy are likely rational and so it is important to consider their agency -- and any impacts -- in taking actions to maximize the value of their holdings.
In short, solutions deployed in infrastructure-centric Blockchain networks must address these aspects in their design when managing token supply.

%or always need to keep a minimum amount of tokens in a central treasury. 
%For example, we might limit the number of tokens minted or burned per day or always need to keep a minimum amount of tokens in a central treasury. 

Our key insight is that token economies can be modeled as dynamical systems, which allows us to leverage powerful ideas from \textbf{mathematical control theory} to maximize a Blockchain network’s utility function under chosen constraints. Control theory is a natural tool since the token economy is a dynamical system – the circulating token supply, token price, and consumer demand change as a function of our burn and mint decisions. Likewise, we have control authority – we are able to adapt the burn or mint mechanisms to regulate the token economy. Moreover, we can design a control cost function that captures key metrics for desired  performance and evolution of the Blockchain dynamical system. Crucially, we can model the dynamics of the system, since we engineer the Blockchain protocol and token economy dynamics. As such, regulating the Blockchain token economy is a model-based control problem, which can be solved using powerful ideas from nonlinear optimization and optimal control theory. 

Overall, the contributions of this paper are three-fold. 
%To the best of our knowledge, we are the first to apply optimal control theory to design blockchain token economies. 
%We introduce a general dynamical systems theoretic formulation for blockchain token economies that  flexibly model fixed-supply as well as burn-and-mint systems, along with various network cost functions.
To the best of our knowledge, we are the first to apply optimal control theory to Blockchain tokenomics and introduce a general-purpose dynamical systems model that flexibly captures both fixed-supply as well as burn-and-mint systems. 
%along with various network cost functions.
We design a control system for a token economy using nonlinear model predictive control (MPC) methods that are used in high-performance, safety-critical applications like autonomous driving \cite{williams2018information,bhardwaj2022storm}, robotic manipulation, and rocket guidance \cite{agrawal2020learning}. We demonstrate that these methods perform better than common heuristic controllers, such as proportional integral derivative (PID) controllers used by some algorithmic stablecoins. Specifically, we improve on PID by
$2.4 \times$ on simulated timeseries demand patterns and by $2.7 \times$ on real demand patterns from the Helium DeWi Blockchain. 
Finally, we introduce a novel game-theoretic formulation for how owners of tokens and a central reserve strategically interact to maximize network welfare.

\paragraph*{Related Work: }
Generally, prior research on Blockchains as dynamical sytems \cite{zargham2018state,zhang2020modeling,caginalp2018dynamical} focus on miner profitability and on the influence of Block rewards on supply and demand dynamics.
Our work differs from existing literature in that we focus on understanding incentives and equilibrium in infrastructure-centric Blockchain systems. Specifically, in our case, the supply is fully specified by the actions of the controller, while the demand is specified via forecasts. Thus, our controller specification is decoupled from the possibly complex trajectory of the demand, and the strength of our controller's predictions relates with the strength of the forecasts used in the system.

In order to better understand the robustness of our methodology, we also consider the impact of rational behavior on the part of the consumers in our system.
To achieve this, game theoretic analyses in Blockchain systems have been used over many years, starting with the original Bitcoin whitepaper \cite{nakamoto2008bitcoin}.
Since the discovery of the selfish mining attack \cite{eyal2018majority}, game theoretic methods have been used to investigate rational deviations \cite{carlsten2016instability}, mining pools \cite{eyal2015miner}, and more recently transaction fee auctions in Ethereum like Blockchains \cite{roughgarden2021transaction,chung2021foundations}.
 Our work differs from existing literature as we focus on the effects of rational behavior on buy-back and pay strategies used to stabilize token prices, and not necessarily on modeling the effects on an underlying Blockchain protocol.

Finally, as our aim is to stabilize a token price in a Blockchain network, our work bears a degree of similarity to algorithmic stable-coins.
However, our interests are in intelligently controlling the circulating supply of a token to balance payments to service providers needed to scale a network with a pre-specified control trajectory on the token price. Thus, our work is more related to service networks employing burn and mint systems such as Helium \cite{HeliumWhitePaper} (which inspired our model) and Factom \cite{snow2014factom} than more general purpose stable-coins like Reflexer \cite{RAI} or Terra \cite{kereiakes2019terra}.
Furthermore, most existing literature is {\em reactive} through the use of heuristic methods such as PID, whereas our work is {\em predictive} through optimal adaptive control methods.
As our focus is on infrastructure networks, our work is applicable  to  DeWi \cite{messie2019baladin} scenarios like Helium \cite{HeliumWhitePaper}, as well as file sharing \cite{FilecoinWhitePaper} and decentralized video streaming \cite{petkanics2018protocol}.

\section{A Primer on Optimal Control}\label{sec:controller}
We now provide a basic primer on optimal control theory, which enables us to naturally model the token economy as a controlled dynamical system. Using this, we describe the state of a dynamical system, the control inputs, dynamics, and the high-level performance criterion (cost function).  

%We now introduce several state-of-the-art control methodologies and how they can be used for blockchain tokenomics. 
The state vector is denoted by $x_t \in \reals^n$, the control vector by $u_t \in \reals^m$, and the dynamics are given by $x_{t+1} = f(x_t, u_t, s_t)$, where $s_t$ is an exogenous timeseries input, such as a demand forecast. Since we have forecasts of node and consumer growth for $H$ steps in the future, we naturally have a finite horizon control problem of $H$ steps. Our
goal is to optimize the performance metric, which is to minimize the aggregate control cost $J$. 
Typically, the control cost $J$ is a sum of a terminal cost and stage costs penalizing state deviations from a reference trajectory (tracking error) and control effort, of the form $J = c_H(x_H) + \sum_{t=0}^{H-1} c_t(x_t, u_t)$, where $c_t(x_t, u_t)$ is a possibly time-variant cost. 

Crucially, we have a good nominal model of the dynamics, since the token economy is under our design. Of course, there are uncertainties which arise due to the stochastic demand forecast $s_t$. Since we have known nominal dynamics, we use standard \textit{model-based} control techniques, which solve an optimization problem to find the optimal set of controls to minimize the cost function subject to dynamics constraints
\cite{Camacho2013,borrelli2017predictive}. 
Moreover, we often constrain the state $x_t$ and control $u_t$ to lie in sets $\mathcal{X}$ and $\mathcal{U}$ respectively to capture, for example, strict actuation limits.
%Moreover, we often have state constraints and control constraints that limit actuation effort, which constrain the state to lie in set $\mathcal{X}$ and controls to lie in set
Thus, the general optimal control problem can be stated as:
%\subsection{Formal Control Problem}\label{subsec:control_problem}
%Now, using Equations (\ref{eq:dynamics}) and (\ref{eq:cost}) we can write finally the formal control problem:
\begin{small}
\begin{equation}\label{eq:formal_problem}
\begin{array}{ll@{}ll}
    \underset{u_{0:H-1}}{\text{minimize}} & \underset{s_{0:H-1}}{\mathbb{E}}\big[J(x_0,u_{0:H-1}; s_{0:H-1})\big], &&\\
    \text{subject to}\;\;&x_{t+1} = f(x_t, u_t, s_t) \quad & \forall t \in \{0, \dots, H-1\} & \\
                         & x_{t} \in \mathcal{X} & \forall t \in \{0, \dots, H\} & \\
                         & u_{t} \in \mathcal{U} & \forall t \in \{0, \dots, H-1\}, & \\ 
\end{array}
\end{equation}
\end{small}
where $x_0$ is the initial state and we plan for $H$ steps given a stochastic demand forecast $s_{0:H-1}$.  
We now model the token economy in Sections \ref{sec:model} and \ref{sec:method}.  
%We now model the token economy as a controlled dynamical system in Section \ref{sec:model}, and propose optimal control strategies in Section \ref{sec:method}. %outline several state-of-the-art optimal control solutions for tokenomics in Section \ref{sec:method}. %by defining our problem's state $x_t$, controls $u_t$, exogenous forecast $s_t$, dynamics $f(x_t, u_t, s_t)$, and cost function $J$. Then, we outline several state-of-the-art optimal control methods as they apply to tokenomics in Section \ref{sec:method}.  %We now model the token economy as a controlled dynamical system by defining our problem's state $x_t$, controls $u_t$, exogenous forecast $s_t$, dynamics $f(x_t, u_t, s_t)$, and cost function $J$. Then, we outline several state-of-the-art optimal control methods as they apply to tokenomics in Section \ref{sec:method}. 

\section{The Token Economy as a Dynamical System}\label{sec:model}
%We now model the token economy for infrastructure networks as a controlled dynamical system. 
%Specifically, we develop the quantities that will serve as the state variables $x_t$, control variables $u_t$, as well as sensible cost functions $J$ for our controller.
To model the token economy for infrastructure networks as a controlled dynamical system, we define our problem's state $x_t$, controls $u_t$, exogenous forecast $s_t$, dynamics $f(x_t, u_t, s_t)$, and cost function $J$. For the dynamics, we show the relationships between the token's circulating supply, price, and reserve used by the controller.
As shown in Fig. \ref{fig:tokenomics}, these quantities will form our state variables which are controlled via buying back tokens and paying tokens to the user base.
%Then, in Section \ref{subsec:control_problem}, we pose our formal control problem. 
%fully make this connection by posing our formal control problem, which we examine and relax in subsequent sections.

%\subsection{The Token Economy as a Dynamical System}\label{subsec:token_dynamics}
We capture the following interaction: {\em Nodes} provide services (5G base stations, EV charging, etc.) for {\em consumers}.
The consumers pay dollars to the controller to use these services, and the controller converts these dollars into tokens at market price to pay to the nodes as rewards for their services.
This reward payment is the {\em minting} action.
Furthermore, the dollars received from consumers is {\em income} that is placed in a dollar reserve.
Funds from this reserve are in turn used to buy back tokens from the nodes at a price posted by the controller.
The bought back tokens are removed from the circulating supply and placed into their own reserve, which is effectively the {\em burn} action. We model the token economy as a closed system, meaning that the power to change the circulating supply is endowed only in the buy back and pay mechanism and no exogenous actor. 
Effectively, this means that no node will intentionally burn their own tokens and lose assets.
Finally, we examine the token economy as a discrete time system.

%\subsubsection{Logically ``Centralized'' Controller  via Distributed Smart Contracts}
%\textbf{Logically ``Centralized'' Controller  via Distributed Smart Contracts}
Before proceeding, one more comment:
Recall, as in Fig. \ref{fig:tokenomics}, the Blockchain network has a reserve that mints new tokens and buys back tokens from the market. Our controller resides in this reserve, and can be considered as logically ``centralized''. Of course, the control logic is implemented by \textit{distributed} nodes running smart contracts. However, all the smart contracts come to a consensus on the key state variables, which are aggregate (not node-specific) quantities such as the circulating supply, token price, total consumer demand
etc. As such, the nodes can decide how much to pay/buy-back in a decentralized way. 
%as a virtual ``central'' controller.

\subsubsection{Circulating Supply}
%\textbf{Circulating Supply}

The circulating supply of tokens $S_t$ increases when $\paytokens$ tokens are paid as rewards to nodes and decreases when tokens are bought back:\vspace{-0.7em}
\begin{small}
\begin{equation}\label{eq:supply}
    S_{t+1} = \totalsupply + \underbrace{\paytokens}_{\text{Tokens Paid}} - \underbrace{\buybacktokens}_{\text{Tokens Bought Back}}.
\end{equation} 
\end{small}

%Here, the units for $\paytokens$ and $\buybackdollars$ are in tokens and dollars respectively.
%Specifically, while $\paytokens$ is the number of tokens paid out to the service providers from the reserve, $\buybackdollars$ is the number of dollars the controller pays to the users to purchase their tokens.
Here, $\buybackdollars$ is the number of dollars the controller pays to active token owners to purchase their tokens.
This means the number of tokens bought back in time $t$ is $\buybackdollars$ divided by the purchasing price $(\tokenprice + \incentive)$.
We define $\tokenprice$ as the current market price of the token, while $\incentive$ is the extra amount the controller pays the users over the market price so as to incentivize them to part with their tokens.
If the token owners are not rational, $\incentive$ simply equals zero.
However, a rational owner may not part with their tokens if they believe the price will increase in successive timesteps, and so the controller may need to pay an added incentive price $\incentive$ to overcome the agents' beliefs.
We formulate a mathematical game to calculate $\incentive$ for rational agents in Section \ref{sec:game}.

%In the presence of rational agents, $\incentive$ can be calculated via a game-theoretic approach, as described in Section \ref{sec:game}.
%by examining an associated simultaneous game.
%We more fully examine the details of this in the sequel.

\subsubsection{Reserve}
%\textbf{Reserve}

The controller will hold two reserves, one being comprised of tokens and the other dollars. 
At time $t$, the values of these reserves are given by $\tokenreserve$ and $\dollarreserve$ respectively.
The dollar reserve increases with the income received from users purchasing tokens, and decreases with buy backs from the market: 
\begin{small}
\begin{equation}\label{eq:reserve_dollars}
    R_{t+1}^{\text{USD}} = \dollarreserve - \buybackdollars + \reserveincomes,
\end{equation}
\end{small}
where $\buybackdollars$ is the dollars used to buy back tokens at time $t$, while the new variable $\reserveincomes$ is the income received from users purchasing tokens. $\reserveincomes$ is a linear function of the number of consumers in the system, since the consumers pay a fixed dollar amount for a unit of infrastructure service. For example, in DeWi networks, a consumer pays a fixed number of USD for each gigabyte of data
\cite{HeliumWhitePaper}. Analagous to the dollar reserve, the token reserve $\tokenreserve$ increases with buy backs and decreases with payments made to the service providers, i.e.,
\begin{small}
\begin{equation}\label{eq:reserve_tokens}
    R_{t+1}^{\text{Token}} = \tokenreserve - \paytokens + \buybacktokens. 
\end{equation}
\end{small}
\subsubsection{Token Price}
The token price at time $t$ is given by $\tokenprice$, and we assume it is market clearing. Thus, for supply $\totalsupply$ and demand $\demand$ at time $t$, we have:
\begin{small}
\begin{equation*}
    \tokenprice = \frac{\demand}{\totalsupply}.
\end{equation*}
\end{small}
The demand for tokens is an (unspecified) function of the number of consumers in the system $\numconsumer$ and the price of the token $\tokenprice$. We  assume that for a fixed price, demand is proportional to the number of consumers, and therefore for a fixed token price, $\reserveincomes \propto \demand$. In practice,  demand can be forecasted using historical data, as shown for public Helium DeWi data in Section \ref{sec:experiments}.

\begin{remark}
\label{remark:vanilla}
Vanilla buy back and pay strategies, like those in many operational BME protocols, reach an equilibrium.
This is a strategy entirely clearing its income in each step, i.e.,
\begin{small}
\begin{align*} 
    \reserveincomes = \buybackdollars, && \paytokens = \buybacktokens.
\end{align*}
\end{small}
Then, by verifying Eqs. \ref{eq:supply}--\ref{eq:reserve_tokens}, one sees that the circulating supply remains invariant, $\totalsupply = S_{t-1} = \ldots = S_0$, as do the reserves $\dollarreserve$ and $\tokenreserve$. 
    %will remain at their initial value as well.
\end{remark}

\subsubsection{Why Control?}
While the simple strategies in Remark \ref{remark:vanilla} are stable, they do not provide much-needed flexibility for tokenomics. For example, we might need to increase the token dollar reserve to pay for upgrades, which requires saving income instead of directly paying it all to nodes. Further, we might want the token price to steadily grow at a certain rate, so that nodes can sell tokens to recover their capital investment costs (e.g.,
buying a 5G hotspot). In essence, we need control to flexibly steer the system away from the equilibrium in Remark \ref{remark:vanilla} to achieve a high-level performance criterion, as formalized next. 
\vspace{-1\baselineskip}
\subsubsection{State, Control and Cost Variables:}
%As mentioned in Section \ref{sec:controller}, we refer to the state and control variables symbollically as $x_t$ and $u_t$ respectively. Informed by the previous discussions, these variables are given by
The state $x_t$ captures the dynamic quantities that are necessary to control the system. Likewise, the control vector $u_t$ consists of how much we adaptively pay, buy-back, and our incentive price: 
\begin{small}
\begin{align*}
    \state{t} = \begin{bmatrix}
            \totalsupply, & \dollarreserve, & \tokenreserve, & \tokenprice
        \end{bmatrix}^{\top}, &&
    \control{t} = \begin{bmatrix}
            \buybackdollars, & \paytokens, & \incentive \end{bmatrix}^{\top}.
\end{align*}
\end{small}
Additionally, as our method is {\em predictive}, we use forecasts predicting the future income and consumer demand. In practice, these can come from data-driven modeling using historical transaction data. 
%state of the system.
%In particular, we assume access to forecasts on the future demand and income which could be obtained via estimation, crowdsourcing, or other techniques.
The forecasts at time $t$ are: 
\begin{small}
\begin{equation*}
    \forecasted{t} = \begin{bmatrix}
            \widehat{\demand}, & \widehat{\reserveincomes}
        \end{bmatrix}^\top.
\end{equation*}
\end{small}
%Now we can fully write the nonlinear dynamics of our system, captured by $f:\state{t},\control{t},\forecasted{t}\mapsto \state{t+1}$, using Equations (\ref{eq:supply}--\ref{eq:reserve_tokens}):
Now, we can write the nonlinear dynamics of our system using Eqs. \ref{eq:supply}--\ref{eq:reserve_tokens}:
\begin{small}
\begin{equation}\label{eq:dynamics}
    \state{t+1} = f(\state{t}, \control{t}, \forecasted{t}) =
    \begin{bmatrix}
        x_{t}(0) + u_{t}(1) - \frac{u_t(0)}{x_t(3)+u_t(2)} \\
        x_{t}(1) + s_{t}(1) - u_{t}(0) \\
        x_{t}(2) + \frac{u_t(0)}{x_t(3)+u_t(2)} - u_{t}(1) \\
        \frac{s_{t+1}(0)}{x_{t+1}(0)}  
    \end{bmatrix}.
\end{equation}
\end{small}
\vspace{-\baselineskip}
\subsubsection{Cost Function}
The cost function $J$ encodes the  token economy's high-level performance criterion. For example, we often want the price to follow a smooth, steadily increasing reference trajectory denoted by $\bartokenprice$. Thus, we penalize the $L_2$ norm difference between the real price $\tokenprice$ and the desired reference price $\bartokenprice$. Likewise, we might want to penalize the enacted controls $u_t$ from a nominal, reference set of controls $\bar{u}_t$. In
practice, these reference controls can come from the income-clearing strategy in Remark \ref{remark:vanilla} or be computed to achieve the reference price trajectory in the absence of forecasting errors. 
Thus, our control cost function is:  
%Thus, our control cost function trades off how closely we achieve the price reference and our extra control effort as follows: 
%We let the cost function be given by the sum of the square errors between reference trajectories for state $\tokenprice$ and control variables $\buybackdollars$ , $\paytokens$. This is,
\begin{small}
\begin{equation}\label{eq:cost}
    J(x_0,u_{0:H-1}) = \mathlarger{\mathlarger{\sum}}_{t=0}^{H}
\begin{bmatrix}
    \beta_1 \\ \beta_2 \\ \beta_3
\end{bmatrix}^\top
\begin{bmatrix}
    \left( x_t(3) - \bar{x}_t(3) \right)^2 \\
    \left( u_t(0) - \bar{u}_t(0) \right)^2 \\
    \left( u_{t}(1) - \bar{u}_t(1) \right)^2
\end{bmatrix}.
%        \beta_1 \left( x_t(3) - \bar{x}_t(3) \right)^2 + \beta_2 \left( u_t(0) - \bar{u}_t(0) \right)^2 + \beta_3 \left( u_{t}(1) - \bar{u}_t(1) \right)^2,
\end{equation}
\end{small}
 Blockchain designers can flexibly set parameters $\beta_1, \beta_2, \beta_3$ to trade off how closely we track the reference price and our control effort. Our formulation is extremely general, since we can easily follow a reference token supply that scales with the number of nodes, or even a fixed reference token supply. 
Indeed, our formulation encompasses fixed supply systems, since we can implement an upper bound on the number of tokens (state constraints). 
Likewise, we can use any differentiable, non-convex cost function amenable to gradient-based optimization. 
%where the social planner's choice of $\beta_1, \beta_2$, and $\beta_3$ control the tradeoffs between prioritizing one reference trajectory over others.

\subsubsection{State/Control Constraints}
We now define the feasible state set $\mathcal{X}$ and control set $\mathcal{U}$.
First, each element in the state $x_t$ should be non-negative and ideally above a safety margin, since they represent the circulating supply, token dollar reserves etc. For example, we might want the dollar reserve above a positive safety margin, specifically $\dollarreserve \ge \dollarreservemin$. 
%to guard against adverse financial scenarios. 
Likewise, we want to pay and buy back a positive number of tokens, given by $\paytokens, \buybackdollars > 0$. 
Finally, we have no constraints on $\Delta p_t$, since we can offer to buy back tokens below market price.

\subsubsection{Formal Control Problem}\label{subsec:control_problem}
Equations \ref{eq:dynamics} -  \ref{eq:cost} define the state space, controls, dynamics, cost function, and constraints for control of the token economy. Thus, we define the formal control problem exactly as in Eq. \ref{eq:formal_problem}. 
Our problem has nonlinear dynamics (since we divide by the price variable $\tokenprice$) and a quadratic cost function. Thus, it is a non-convex optimization problem, which can be solved using the nonlinear optimal control methods introduced next.

\section{Control Design Methodology}
\label{sec:method}

%We now illustrate how to solve our formal control problem Eq. \ref{eq:formal_problem}, where the state, dynamics, and cost function are defined in Eqs. \ref{eq:dynamics}-\ref{eq:cost}. To do so, we first outline powerful methods from nonlinear optimal control theory.

We now illustrate methodologies that solve our formal control problem Eq. \ref{eq:formal_problem} by outlining  techniques from nonlinear optimal control theory.

%We observe that the dynamics are differentible (smooth), but nonlinear, due to division by the token price. Moreover, the cost is smooth and quadratic. Thus, we can invoke powerful methods from nonlinear optimal control theory, as outlined below. 
%\textcolor{red}{showing how to relax problem (\ref{eq:formal_problem}) and solve with iLQR/SCP should go here.}

\textbf{Iterative Linear Quadratic Regulator (iLQR): } 
A canonical model in optimal control is the linear quadratic regulator (LQR), where we have linear dynamics and a quadratic cost function that weights state tracking error and control effort, such as energy expenditure \cite{aastrom2021feedback}. The linear dynamics are given by $x_{t+1} = A_t x_t + B_t u_t$ and the cost function is given by $J(x_0, u_{0:H-1}) = x_H^T Q_H x_H + \sum_{t=0}^{H-1} x_t^T Q_t x_t + u_t^T R_t u_t$, where all $Q_t, R_t$ are positive (semi)-definite matrices  to ensure non-negative costs. 
The optimal solution to LQR is an affine state feedback controller of the form $u^{*}_t = K_t x_t + k_t$, where $K_t$ and $k_t$ are obtained via dynamic programming and the discrete Algebraic Ricatti Equation. 
%Alternatively, LQR can be solved as a convex quadratic program (QP) \cite{boyd2004convex}.

In practice, we desire that our system follows a non-zero reference trajectory given by ${(\bar{x}_t, \bar{u}_t)}_{t=0}^H$, such as regulating the system to track a time-varying token supply. Moreover, the dynamics $f_t(x_t, u_t)$ could be non-linear and the cost function $c_t(x_t, u_t)$ could be non-convex nor quadratic. The key insight behind \textit{iterative} LQR (iLQR) \cite{van2014iterated} is that we \textit{linearize} a system's nonlinear dynamics around the reference trajectory $(\bar{x}_t,
\bar{u}_t)$ if the dynamics are differentiable. Thus, we compute the Jacobian matrix of the nonlinear dynamics anchored at the current reference $\bar{x}_t, \bar{u}_t$ to obtain the dynamics matrix $A_t = \frac{\partial f_t}{\partial x}(\bar{x}_t, \bar{u}_t)$ and control matrix $B_t = \frac{\partial f_t}{\partial u}(\bar{x}_t, \bar{u}_t)$. Likewise, for a twice continuously differentiable cost function, we use the Taylor Series to form its local quadratic approximation. 

Given a nominal reference trajectory, linearized dynamics, and a quadratic approximation of the cost, we can simply invoke LQR to improve our nominal reference trajectory. The process repeats until the control cost converges, which is analagous to Newton's method. We can incorporate strict state or control constraints by adding them as penalties to iLQR's cost function using \textit{Augmented Lagrangian} iLQR (AL-iLQR) methods. 
Crucially, AL-iLQR is our solution method of choice for tokenomics, since we have smooth nonlinear dynamics, a quadratic cost function, a well defined reference trajectory for the token price/circulating
supply, and strict constraints for non-negative treasuries.

\textbf{Sequential Convex Programming (SCP): }
SCP extends the core ideas behind iLQR to control problems with strict state or control constraints \cite{bonalli2019gusto,malyuta2021convex}. First, we linearize the dynamics around a reference trajectory, just as in iLQR. Then, we form a convex approximation of the cost function, often using a local quadratic approximation. Thus, we recover a constrained convex optimization problem, which we solve to obtain a new nominal trajectory.
We then re-linearize around the updated nominal trajectory until the control cost saturates.

%We then find a new convex relaxation of the original problem around the updated nominal trajectory and repeat until the control cost saturates.
%For more details on iLQR and SCP, we refer the reader to \cite{malyuta2021convex,tassa2012synthesis}. 

\textbf{Model Predictive Control (MPC): }
In practice, we receive a stochastic forecast $s_t$ of income from consumer demand. Then, given the current state $x_0$, and forecast $\hat{s}_{0:H-1}$, we can propagate the known dynamics $f(x_t, u_t, s_t)$ and solve the control problem in Eq. \ref{eq:formal_problem} to obtain an optimal sequence of controls $u^{*}_{0:H-1}$. In MPC, we simply implement the first control $u^{*}_{0}$, observe the next state $x_1$, and re-plan an optimal trajectory to
mitigate forecast uncertainty \cite{borrelli2017predictive}. 
%The name MPC is natural since we use a dynamics model, predict the system evolution, and optimize a cost function for control. 
Crucially, all our experiments run MPC where the underlying control strategy in Eq. \ref{eq:formal_problem} can use AL-iLQR or SCP without loss of generality.

\textbf{Proportional Integral Derivative (PID) Control: }
Our proposed solution uses AL-iLQR and SCP since we can model the token dynamics and explicitly desire to optimize a cost function. We compare these methods to a benchmark proportional integral derivative (PID) controller. While PID controllers achieve stability \cite{aastrom2021feedback}, they do not explicitly optimize a cost function like iLQR/SCP, and often require extensive tuning and can overshoot a reference trajectory. PID is a fitting
benchmark due to its simplicity and the fact that recent algorithmic stablecoins, such as the RAI index \cite{RAI}, use PID.

A PID controller operates on an error signal, which penalizes the deviation of the current state and desired/reference state. In our setting, the error signal $e_t = \bartokenprice - \tokenprice$ is the difference between the reference token price $\bartokenprice$ and actual, measured token price $\tokenprice$. Then,  a PID controller generates a control decision $u_t$ that is proportional to the instantaneous error, integrates past errors, and forecasts how the error will change as: $u_t = k_p e_t + k_i \sum_{j=0}^{t-1} e_j + k_d e_{t+1}$. Here, $k_p, k_i,k_d$ are gain parameters that are often manually tuned for stability.

%$u_t = k_p (S^{\mathrm{ref}}_t - S_t) + k_i \sum_{k=0}^{t-1} (S^{\mathrm{ref}}_t - S_t) + k_d (S^{\mathrm{ref}}_{t+1} - S_{t+1})$.

\section{Strategic Pricing: A Game-Theoretic Analysis}\label{sec:game}
\begin{figure}[t]
    \begin{minipage}[c]{0.5\textwidth}
    \includegraphics[width=\columnwidth]{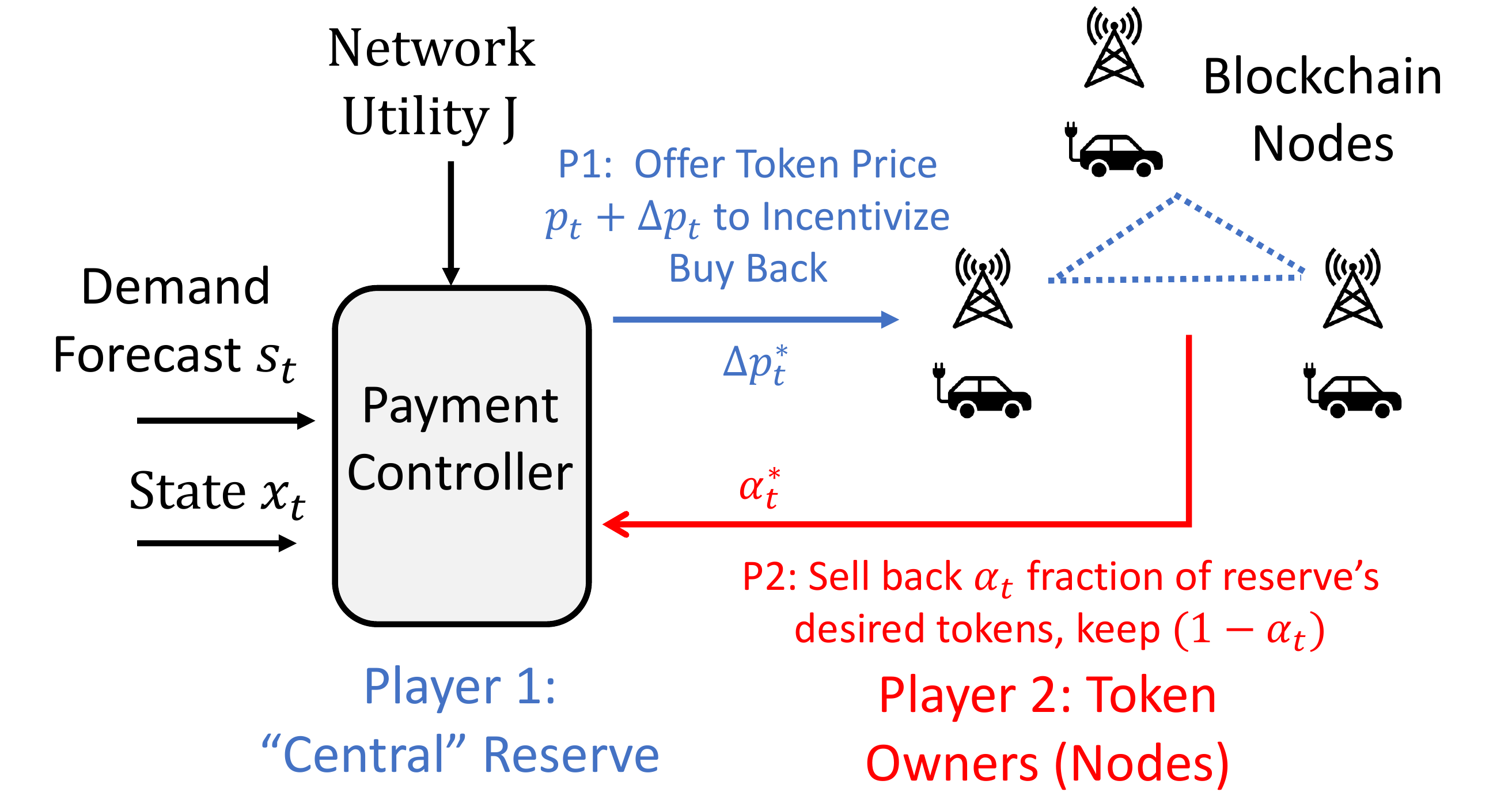}
      \end{minipage}\hfill
      \begin{minipage}[c]{0.48\textwidth}
\caption{\small{\textbf{Strategic Pricing as a Game: } 
    %We formulate a game between the reserve (player 1) and all token-owning nodes (collectively, player 2). 
          The reserve (player 1) offers to buy-back tokens at the market price plus a bonus incentive of $\incentive$ to reduce the circulating supply. However, all token-owning nodes (collectively, player 2) might want to only sell a fraction $\alpha_t$ of tokens since they could be valuable in the future. 
%However, if they refuse to sell back any tokens, the price might plummet due to an excess circulating token supply. 
    %and inflation.
    We solve for an optimal game strategy for each player ($\Delta p_t^*$ and $\alpha^*$) using bi-level optimization.
    \label{fig:game}
} }
    \end{minipage} 
\vspace{-3em}
\end{figure}

In our control-theoretic formulation, we assume token owners will gladly sell their tokens to the reserve when it offers to buy back tokens with an incentive price of $\Delta p_t$. However, as shown in Fig. \ref{fig:game}, strategic token owners might only sell a fraction of their tokens for immediate revenue and retain the rest for their future expected value. As such, the reserve must offer a sufficiently high incentive
$\Delta p_t$ to goad token owners to sell their valuable tokens so that the circulating token supply is regulated to avoid inflation. Our key insight is that strategic pricing can be formulated as a two-player Stackelberg game (see \cite{osborne2004introduction}). 
%that can be solved using bi-level optimization methods. 
%\input{fig_latex/fig2_game}

\subsubsection{Market Dynamics}
%\label{subsec:market_dynamics}
The market dynamics arise from the selfish behavior of rational consumers.
At each timestep, we have a two-step sequential game of complete information between the reserve's controller (player 1)  and all token-owning nodes (player 2). The controller optimizes program (\ref{eq:formal_problem}) and the consumers seek to maximize the value of their token holdings over the time horizon.
By complete information, we mean that facts about the opponent respectively are common knowledge.
For example, token owners are aware of the controller's strategy, which is encoded in smart contracts distributed across the blockchain. Likewise, each player can also perfectly observe the token price and supply. 
%is perfect and common knowledge.

The controller moves first by posting a price $(\tokenprice + \incentive)$.
Then, the consumers respond by selling an $\alpha_t$-fraction of their collective holdings, where $\alpha_t$ is chosen to maximize the value of the consumers' holdings (utility $U_t$) at timestep $t$. 
Specifically, the utility $U_t$ is a sum of the price they can get {\em today} and the price they can get {\em tomorrow}, which is a function of the game strategies:
\begin{small}
\begin{align*}
    \utility = \underbrace{\alpha_t \cdot\totalsupply(\tokenprice+\incentive)}_{\substack{\text{Current earnings from selling} \\ \text{ $\alpha_t$-fraction of supply.}}}  + \underbrace{(1-\alpha_t) \cdot \gamma \totalsupply \mathbb{E}(p_{t+1}^{\text{Tok}})}_{\substack{\text{Future expected earnings from} \\ \text{holding $\alpha_t$-fraction of supply.}}}.
\end{align*}
\end{small}

In the above, $\gamma$ is a risk factor attenuating the expected future earnings from not selling. Further, the randomness in the expectation is due to forecasting noise as our controller is not randomized. Thus, for any controller-chosen incentive $\incentive$, the nodes' optimal strategy is to choose $\alpha_t$ such that: 
%So, in the game the optimal consumer strategy, which is to say their choice of $\alpha_t$, is given by the following:
\begin{small}
\begin{align}\label{eq:consumer_strategy}
    \max_{\alpha_t \in [0,1]}U_t(\alpha_t, \incentive) \quad \quad \textsc{Node Utility Optimization}.
\end{align}
\end{small}
Moreover, due to program (\ref{eq:consumer_strategy}), the controller can compute the consumer strategy for any incentive price $\incentive$.
%This means that we can transition the control problem (\ref{eq:formal_problem}) into the setting with incentives by constraining the amount of tokens the controller can buy back by a function of the consumer strategy,
This means that we can transition the control problem (\ref{eq:formal_problem}) into the setting with incentives by equating the amount of tokens the controller buys back to the amount of tokens the nodes agree to sell:
%This means that we can transition the control problem (\ref{eq:formal_problem}) into the setting with incentives by constraining the amount of tokens the controller can buy back by a function of the consumer strategy,
\begin{small}
    \begin{align}\label{eq:extra_constraint}
    \buybacktokens = \alpha_t \cdot \totalsupply,
\end{align}
\end{small}
where $\alpha_t$ is computed as a function of incentive price through program (\ref{eq:consumer_strategy}).
This means that in practice the controller can account for selfish behavior through roll out simulations to identify good incentive prices.
However, we now formalize how the controller can play optimally by identifying the incentive price $\incentive^*$ inducing a subgame perfect equilibrium $(\alpha_t^*, \incentive^*)$.
%This makes a bilevel optimization problem with which our framework affords us the ability to make strong analytical statements.
%This is, in the sequel we examine the existence of Nash equilibria in this game, as well as the {\em price of anarchy} which bounds the largest gap between the objective value the controller obtains in the setting with and without incentives.
%However, what would be even better is if the controller could play optimally by identifying the incentive price $\incentive^*$ inducing the Nash equilibrium maximizing program (\ref{eq:formal_problem}), $(\alpha_t^*, \incentive^*)$.
%This makes a bilevel optimization problem with which our framework affords us the ability to make strong analytical statements.
%This is, in the sequel we examine the existence of Nash equilibria in this game, as well as the {\em price of anarchy} which bounds the largest gap between the objective value the controller obtains in the setting with and without incentives.

\subsubsection{A Stackelberg Game for Strategic Pricing}
%\label{subsec:stackelberg}
%We formalize a game between the controller and nodes. First, the central reserve offers an incentive price $\Delta p_t$. Then, as a function of $\Delta p_t$, the token owners make a collective decision to only sell a fraction $\alpha_t$ of their tokens. 
Since the controller first posts a price $\incentive$ and the nodes respond with the fraction $\alpha_t$ of the holdings they wish to sell, we naturally have a leader-follower (Stackelberg) game. As mentioned above, we use (\ref{eq:extra_constraint}) to constrain the tokens bought back by $\alpha_t$. Then, recalling that the node's strategy is given by (\ref{eq:consumer_strategy}), the controller's optimization problem is: 

\begin{minipage}[t]{0.50\textwidth}
    \parbox[t][3.5cm][t]{\linewidth}{
          \begin{subequations} \label{eq:reservegame}
          \centering
          \textsc{Game Formulation}
          \begin{small}
          \begin{align}
            \underset{u_{0:H-1}}{\min} & \underset{s_{0:H-1}}{\mathbb{E}}\big[J(x_0,u_{0:H-1}; s_{0:H-1})\big], && \notag\\
            \text{s.t,}\;\;&x_{t+1} = f(x_t, u_t, s_t) & & \notag\\
                                 & x_{t} \in \mathcal{X}, u_{t} \in \mathcal{U}  & & \notag\\
                                 & \buybackdollars = \alpha_t \totalsupply (\tokenprice + \incentive) && \notag\\
                                 & \boxed{\alpha_t = \arg\,\max_{\alpha_t \in [0,1]} U_t(\alpha_t, \incentive)} && \notag
          \end{align} 
          \end{small}
          \end{subequations}
    }
   \end{minipage}%
  \hfill
   \begin{minipage}[t]{0.50\textwidth}
    \parbox[t][3.5cm][t]{\linewidth}{
      \begin{subequations} \label{eq:tokenownergame}
      \centering
      \textsc{Bi-Level Optimization}
          \begin{small}
          \begin{align}
            \underset{\substack{u_{0:H-1}, \lambda_{1:2}\\ \alpha_{0:H-1}}}{\min} & \underset{s_{0:H-1}}{\mathbb{E}}\big[J(x_0,u_{0:H-1}; s_{0:H-1})\big], && \notag\\
            \text{s.t,}\;\;&x_{t+1} = f(x_t, u_t, s_t) & & \notag\\
                                 & x_{t} \in \mathcal{X}, u_{t} \in \mathcal{U}  & & \notag\\
                                 & \buybackdollars = \alpha_t \totalsupply (\tokenprice + \incentive) && \notag\\
                                 & \color{blue} \boxed{ \nabla_{\alpha}\mathcal{L}(\alpha_t,\incentive,\lambda) = 0} && \notag \\
                                 & \color{blue}\boxed{ \lambda_i \leq 0, \alpha_t \in [0,1]} && \notag \\
                                 & \color{blue}\boxed{ \lambda_1 \alpha_t =0, \lambda_2 (1 - \alpha_t) = 0} && \notag 
          \end{align} 
          \end{small}
  \end{subequations}
    }
   \end{minipage}%
   \vspace{2\baselineskip}

%\begin{minipage}[t]{0.50\textwidth}
%    \parbox[t][3.5cm][t]{\linewidth}{
%          \begin{subequations} \label{eq:reservegame}
%          \centering
%          \textsc{Central Reserve}
%          \begin{small}
%          \begin{align}
%            \underset{u_{0:H-1}}{\min} & \underset{s_{0:H-1}}{\mathbb{E}}\big[J(x_0,u_{0:H-1}; s_{0:H-1})\big], && \notag\\
%            \text{s.t,}\;\;&x_{t+1} = f(x_t, u_t, s_t) & & \notag\\
%                                 & x_{t} \in \mathcal{X} & & \notag\\
%                                 & u_{t} \in \mathcal{U} & & \notag\\
%                                 & \buybackdollars \leq \alpha_t \totalsupply (\tokenprice + \incentive) && \notag\\
%                                 & \alpha_t = \arg\,\min_{\alpha_t \in [0,1]} U_t(\tokenprice, \incentive) && \notag
%          \end{align} 
%          \end{small}
%          \end{subequations}
%    }
%   \end{minipage}%
%  \hfill
%   \begin{minipage}[t]{0.50\textwidth}
%    \parbox[t][3.5cm][t]{\linewidth}{
%      \begin{subequations} \label{eq:tokenownergame}
%      \centering
%      \textsc{Node Token Owners}
%      \begin{small}
%      \begin{align}
%        \underset{\alpha_{t}}{\min} & \utility, &&\notag\\
%        \text{s.s}\;\;&\alpha_t \in [0,1] && \notag
%  \end{align}
%      \end{small}
%  \end{subequations}
%    }
%   \end{minipage}%
%

\subsubsection{Solution via Bi-Level Optimization}
\label{subsec:bilevel}

Notice that the Stackelberg game can be cast as a \textit{bi-level} optimization problem \cite{colson2007overview,bard2013practical}, which is a nested optimization problem where an \textit{outer} optimization problem involves a decision variable that is, in turn, the solution to a second \textit{inner} problem. In our setting, the outer problem is the controller's non-convex control problem (\ref{eq:formal_problem}), which outputs $\Delta
p_t^*$ and requires $\alpha_t^*$ as input. Here, $\alpha_t^*$ is the solution to the inner utility maximization problem of the nodes. Crucially, we observe that the inner utility maximization problem is convex for any given value of $\Delta p_t$, and that Slater's constraint qualification condition holds. Thus, we know that the Karush-Kuhn-Tucker (KKT) conditions are sufficient for optimality for the inner problem. Then, we simply reduce the bi-level (nested)
optimization problem as a single optimization problem for the controller where the optimal choice of $\alpha^*$ is encoded by adding its KKT conditions as extra constraints, with the Lagrangian denoted by $\mathcal{L}$. For clarity, we show the KKT constraints in blue above.

%%Specifically, our optimization problem becomes: 
%\begin{subequations} \label{eq:reservegame}
%  \centering
%  \textsc{Central Reserve}
%  \begin{small}
%  \begin{align}
%    \underset{u_{0:H-1}}{\min} & \underset{s_{0:H-1}}{\mathbb{E}}\big[J(x_0,u_{0:H-1}; s_{0:H-1})\big], && \notag\\
%    \text{s.t,}\;\;&x_{t+1} = f(x_t, u_t, s_t) & & \notag\\
%                         & x_{t} \in \mathcal{X} & & \notag\\
%                         & u_{t} \in \mathcal{U} & & \notag\\
%                         & \buybackdollars \leq \alpha_t \totalsupply (\tokenprice + \incentive) && \notag\\
%                         & \alpha_t = \arg\,\min_{\alpha_t \in [0,1]} U_t(\tokenprice, \incentive) && \notag
%  \end{align} 
%  \end{small}
%  \end{subequations}

\begin{remark}
    \label{remark:nash}
    Since we have a Stackelberg Game, the horizon is finite and a subgame perfect equilibrium can be found via backward induction. First, the best response function of the nodes is calculated. Then, the controller picks an action maximizing its utility, anticipating the follower's best response. For more details, see \cite{osborne2004introduction}. Then it is clear that our method finds a subgame perfect equilibrium since the KKT conditions of the inner problem give a certificate of optimal play on the nodes part. Encoding them as extra constraints on the part of the controller simply gives an explicit route for backward induction in this game.
\end{remark}

%We now show strong experimental performance of our approach in simulations with real DeWi demand data. 

\section{Experiments}\label{sec:experiments}
The goal of our evaluation is to show that (i) using control theory enables us to achieve a more stable, increasing token price and (ii) we can reduce control cost using an optimization-based controller instead of a reactive PID controller. We now describe our benchmark algorithms and evaluation metrics. 

\paragraph*{Evaluation Metrics and Benchmark Algorithms: }

We compare reserve controllers on the following metrics: 

\begin{itemize}[leftmargin=*]
    \item \textbf{Stable Token Price}: We track whether the token price is increasing, its volatility, and the mean squared error (MSE) from a reference price trajectory.
    \item \textbf{Control Cost}: This is a weighted sum of the tracking error (MSE between the token price and reference price) as well as control effort. 
\end{itemize}

We report these metrics for various realistic scenarios where (i) the supply of nodes out-paces the consumer demand, (ii) the supply and demand roughly match, and (iii) the supply lags the consumer demand. 
%Further, we plot a distribution of control cost over many scenarios and initial conditions (starting token supplies) to emulate the controller in various scenarios. 
Our experiments compare the following schemes: 

\begin{itemize}[leftmargin=*]
    \item \textbf{Model Predictive Control (MPC)}: 
We implement the predictive, optimal control scheme proposed in Section \ref{sec:method}.
%, where the controller receives a stochastic forecast $s_t$ of income and consumer demand. Since we have smooth nonlinear dynamics and a quadratic cost function, we run MPC with an underlying SCP or iLQR controller, which work equally well in our experiments. 
        %This is our scheme proposed in Section \ref{sec:method}, where the controller receives a stochastic forecast $s_t$ of income and consumer demand. Since we have smooth nonlinear dynamics and a quadratic cost function, we run MPC with an underlying SCP or iLQR controller, which work equally well in our experiments. 
%        The controller's optimization problem  aims to find a finite sequence of controls to minimize the control cost, which exploits the token economy's model-based dynamics outlined earlier. If we have nonlinear dynamics and a quadratic approximation of control cost, but no strict state or control constraints, we can use the iterative Linear Quadratic Regulator (iLQR) controller. Likewise, if we have strict control constraints, we use Sequential Convex Programming (SCP), specifically Sequential Quadratic Programming (SQP).
	\item \textbf{Proportional Integral Derivative (PID)}: PID controllers are a fitting benchmark since they are widely used in industrial systems such as cruise control, robotic manipulators, and in some algorithmic stablecoins. 
        %However, PID does not explicitly optimize for a high-level cost function like MPC.
    \item \textbf{No Control}: We consider the worst case where the economy has no adaptive control and simply uses the income clearing strategy from Remark \ref{remark:vanilla}. 
\end{itemize}

\subsubsection{Experiment Design}
First, we evaluate our control system on synthetic node supply and consumer demand/income growth data. Then, we repeat our simulations with real-world demand timeseries from the Helium DeWi network. In all experiments, we have a noisy forecast $\hat{s}_t$ of the demand and income growth for a horizon of $H$ future steps.  
Fig. \ref{fig:control_loss} plots a distribution of control cost over many diverse timeseries patterns and initial conditions $x_0$ (initial token supplies etc.) to emulate each controller in various scenarios. 
%We implement the various control benchmarks starting with different initial conditions $x_0$ representing the starting token supply etc. 
Second, we visualize the resulting token supply $S_t$ and token price $\tokenprice$ compared to the desired reference trajectory in Fig. \ref{fig:exp}. All our experiments are coded in Python, run within a few minutes on a standard laptop, and use the Gurobi Optimization package \cite{gurobi2018gurobi}.

%Finally, we report a distribution of control costs across various timeseries patterns in the boxplots in Fig. \ref{fig:control_loss}.
\subsubsection{Experiments with Helium and Synthetic Data}
\begin{figure}[t]
    \includegraphics[width=1\columnwidth]{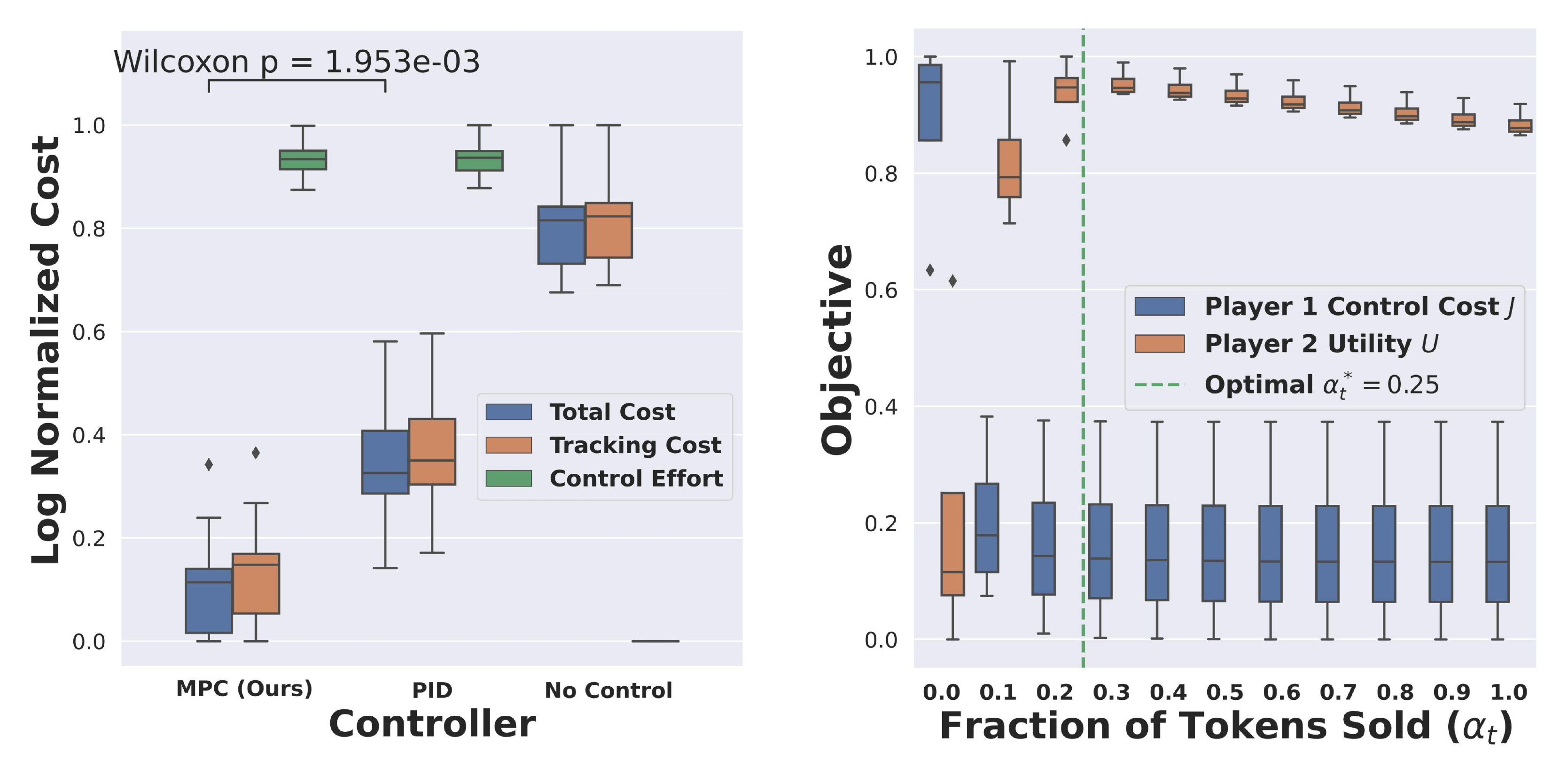}
    \vspace{-2em}
\caption{\small{\textbf{Control Cost with Synthetic Data:}
    (Left)  Our MPC scheme achieves a lower total cost than the PID and the No Control benchmarks by exerting control effort to actively track the price trajectory. (Right) Our bi-level optimization approach finds a game strategy to sell only $\alpha^*_t$ tokens (green line) to maximize node utility and minimize control cost compared to simply holding ($\alpha_t = 0$) or selling all tokens ($\alpha_t = 1$). 
    \label{fig:control_loss}
    \vspace{-1.5em}
} }
\end{figure}
Our synthetic experiments aim to generate realistic supply, demand, and income growth patterns with additive, zero-mean Gaussian noise. For example, we use a sigmoidal growth pattern where demand and supply rapidly rise during the middle of network adoption, but slowly taper off. We also simulate a logarithmic growth function where the demand steadily grows over time. Each simulation trial has a different reference price trajectory to allow a stable price growth, such as a logarithmic
reference. In each trial, the initial token price $\tokenprice$ is randomly chosen away from the reference, and show the controller can dynamically regulate the system. 

\begin{figure}[H]
\centering
  \includegraphics[width=1\columnwidth]{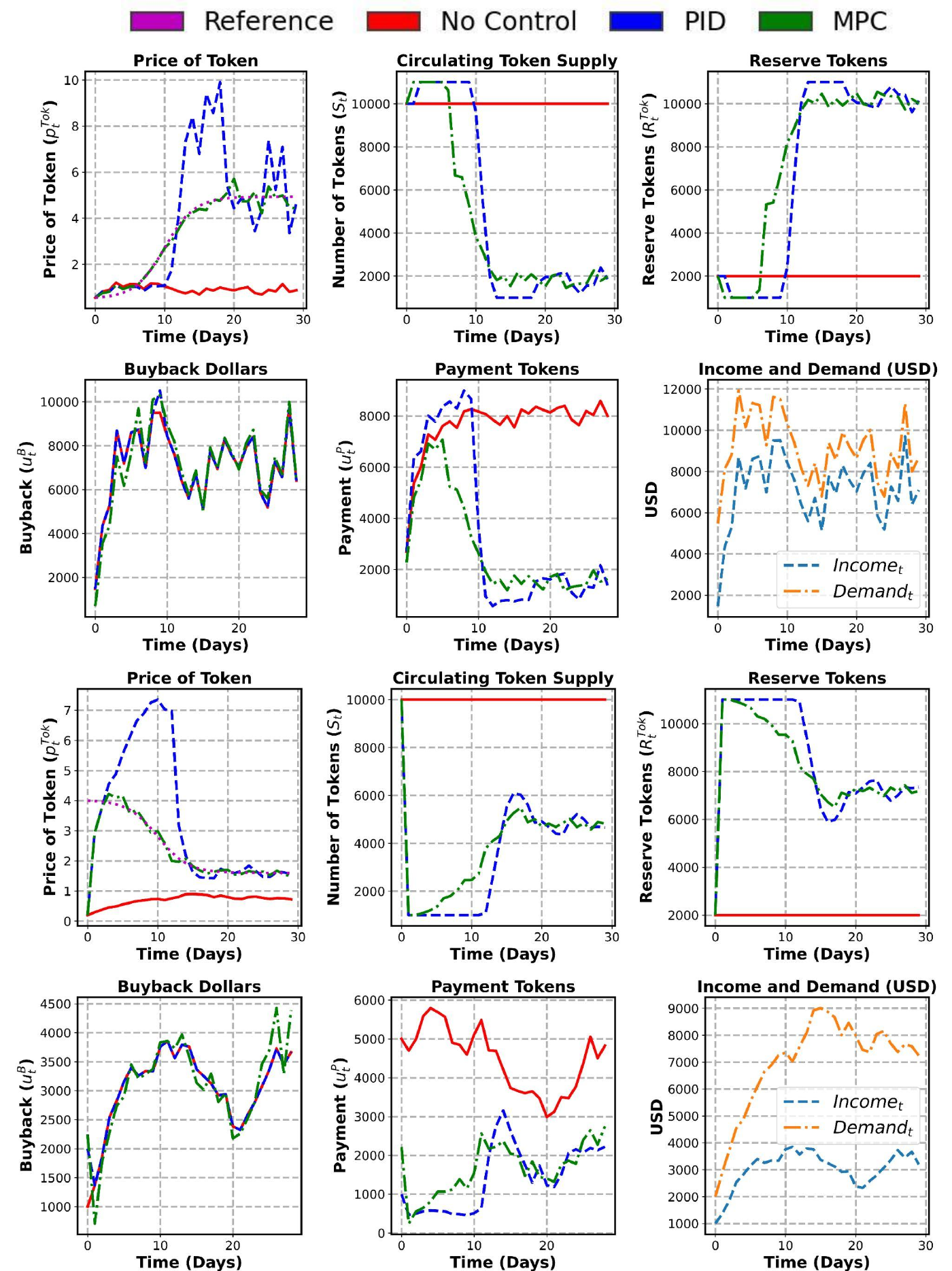}
\caption{\small{\textbf{Results with Helium (top) and Synthetic (bottom) Data:}
    Our key result is on the top left for the token price. Our MPC scheme (green) is able to track the reference (purple) extremely well, while reactive PID (blue) captures the general trend but is highly oscillatory. Crucially, the price plummets without control since we pay too many tokens, which is mitigated by our adaptive MPC payment scheme. The bottom two rows show another price reference to illustrate the generality of our approach.
    %our MPC scheme adaptively curtails token payments to reduce the circulating supply and avoid inflation (middle).
    %We simulate the pricing schemes with Helium data and compare the results between MPC, PID and No Control cases. We observe that MPC is able to achieve a better performance than PID and No Control case.
\label{fig:exp}
\vspace{-1.5em}
} }
\end{figure}

We then repeat the same experimental procedure, but with real timeseries of node growth and consumer demand from the Helium DeWi network, which uses a BME protocol. First, since the Helium growth patterns are smooth, we use a classical Auto-Regressive Integrated Moving Average (ARIMA) forecasting model to predict network growth $H=20$ days in advance. Then, we
implement our controller on a distribution of reference price trajectories. Fig. \ref{fig:exp} shows that MPC significantly reduces the control cost compared to PID benchmarks. 

%Fig. \ref{fig:exp} shows two example trajectories of our system. Our key result is on the top left for the token price. Our MPC scheme (green) is able to track the reference (purple) extremely well, while heuristic PID captures the general trend but is highly oscillatory since it is \textit{reactive}. Crucially, the price plummets without control since we pay too many tokens, which causes inflation. However, our MPC scheme adaptively curtails token payments to reduce the circulating supply and avoid inflation (middle). Importantly, the vanilla income-clearing strategy from Remark \ref{remark:vanilla} (red) immediately pays out the exact same number of tokens it buys back from the market. Thus, the net change in the circulating token supply (and hence reserve) is zero, as indicated by the horizontal red lines for the token plots. 

\paragraph*{Does MPC reduce control cost compared to heuristic controllers?}
We now evaluate our ultimate performance metric, the control cost, across a wide variety of growth patterns and initial conditions. 
Specifically, we used 3 growth patterns with Gaussian noise (sigmoidal, logarithmic, exponential) and many scenarios where demand outstrips supply and vice versa. 
Fig. \ref{fig:control_loss} shows the overall control cost, tracking error, and control effort for all 3 benchmarks. Clearly, MPC achieves a lower control cost than the PID heuristic (Wilcoxon p-value of $.001953$ is statistically significant at the $0.05$ level). As shown in Fig. \ref{fig:control_loss}, the key reason for this difference is that PID is largely \textit{reactive} -- it proportionally responds to the current error and integrates the cumulative error, but does not forecast the future system state accurately to optimize performance. 
In stark contrast, MPC explicitly solves an optimization problem to minimize the control cost. 
%over the forecast horizon $H$. 
%implements the first control, and re-plans to react to estimation errors, which improves ultimate performance.
%In stark contrast, MPC explicitly solves an optimization problem to minimize the control cost over the forecast horizon $H$, implements the first control, and re-plans to react to estimation errors, which improves ultimate performance.

\paragraph*{Does our Stackelberg game solution maximize each player's objective function?}
Fig. \ref{fig:control_loss} (right) illustrates the token owning nodes' utility $U_t$ (player 2) and controller cost function $J$ (player 1). 
Clearly, our bi-level optimization approach is able to find an optimal strategy $\alpha^*_t$ of tokens to sell back, since the node utility is maximized and controller cost is minimized. The solution from bi-level optimization clearly outperforms naive strategies of simply holding all tokens with $\alpha_t = 0$ and selling back all
tokens with $\alpha_t=1$. As expected, the control cost 

\noindent
$J$ (blue boxplots) decreases as $\alpha_t$ increases, since the controller is able to buy back more of the tokens that it desires. Likewise, the node utility (orange) is low when they hold all their tokens ($\alpha_t = 0$) since the price drops because the controller can not remove tokens from circulation through buy backs.

\paragraph*{Does MPC outperform benchmarks to yield a stable token price growth?}
Fig. \ref{fig:exp} shows two example trajectories of our system, where the top two rows are with real Helium data and the bottom two are with synthetic data. Our key result is on the top left for the token price. For the Helium data, our MPC scheme (green) is able to track the reference (purple) extremely well, while heuristic PID captures the general trend but is highly oscillatory since it is \textit{reactive}. Crucially, the price plummets without control since we pay too many tokens, which causes inflation. However, our MPC scheme adaptively curtails token payments to reduce the circulating supply and avoid inflation (middle). Importantly, the vanilla income-clearing strategy from Remark \ref{remark:vanilla} (red) immediately pays out the exact same number of tokens it buys back from the market. Thus, the net change in the circulating token supply (and hence reserve) is zero, as indicated by the horizontal red lines for the token plots. 

The last two rows confirm the generality of our approach -- we can just as easily follow a smoothly decreasing price trajectory.  For example, we might want to price the token more affordably over time. Crucially, the initial token price (row 3, top left) is very low and far from the reference for all schemes, but our MPC method (green) quickly rises to track the reference unlike PID (blue), which overshoots. Importantly, the price without control is very low but slightly increasing since the income/demand gradually increase over time. 
Finally, MPC slowly increases the adaptive payments after timestep 11 for the price to decrease.

\paragraph*{Limitations}
Our trace-driven simulations are limited by offline, historical Helium DeWi data. However, the growth of nodes and consumers might significantly deviate from historical patterns if we actually implemented our proposed controller in the network. In future work, we plan to answer such ``what-if'' questions using recent advances in counterfactual analysis \cite{morgan2015counterfactuals,pearl2010causal,agarwal2020synthetic}. 
%Further, we are working with infrastructure blockchains to deploy our control strategies. 

\section{Discussion and Future Work}\label{sec:discussion}
Our central thesis is that Blockchain tokenomics should be programmable
and dynamically adapt to node growth and consumer demand. Our key contribution is to model
a token economy as a controlled dynamical system, which allows us to leverage rigorous systems theory to design token economies that meet high-level performance metrics (network cost functions). 
We believe our work is timely as several Blockchain projects are working with burn and mint strategies and our  framework enables us to (a) explicitly prove these systems reach a stable equilibrium and (b) flexibly steer this equilibrium to incentivize stable network growth. We are working with several Blockchain projects to instantiate these ideas in practice, which we hope to report on in future work.


%\printbibliography
\newpage
\bibliographystyle{splncs04}
\bibliography{ref/full}

%\end{thebibliography}
\end{document}